\documentclass[aps,prl,twocolumn,superscriptaddress]{revtex4-1}

\usepackage{graphicx}
\usepackage{amssymb}
\usepackage{amsmath}
\usepackage{epstopdf}
\usepackage{hyperref}
\usepackage{subfigure}
\usepackage{braket}
\usepackage{xcolor}

\newcommand*\diff{\mathop{}\!\mathrm{d}}

\begin{document}


\title{Observation of Interaction of Spin and Intrinsic Orbital Angular Momentum of Light}


\author{Dashiell L. P. Vitullo}
\affiliation{Department of Physics and Oregon Center for Optical, Molecular, \& Quantum Science, University of Oregon, Eugene, OR 97403, USA}
\email[]{raymer@uoregon.edu}

\author{Cody C. Leary}
\affiliation{Department of Physics, College of Wooster,  Wooster, Ohio 44691, USA}
\author{Patrick Gregg}
\affiliation{Department of Electrical \& Computer Engineering, Boston University, Boston, MA 02215, USA}
\author{Roger A. Smith}
\author{Dileep V. Reddy}
\affiliation{Department of Physics and Oregon Center for Optical, Molecular, \& Quantum Science, University of Oregon, Eugene, OR 97403, USA}
\author{Siddharth Ramachandran}
\affiliation{Department of Electrical \& Computer Engineering, Boston University, Boston, MA 02215, USA}
\author{Michael G. Raymer}
\affiliation{Department of Physics and Oregon Center for Optical, Molecular, \& Quantum Science, University of Oregon, Eugene, OR 97403, USA}


\date{\today}

\begin{abstract}
Interaction of spin and intrinsic orbital angular momentum of light is observed, as evidenced by length-dependent rotations of both spatial patterns and optical polarization in a cylindrically-symmetric isotropic optical fiber. Such rotations occur in straight few-mode fiber when superpositions of two modes with parallel and anti-parallel orientation of spin and intrinsic orbital angular momentum (IOAM=$2\hslash$) are excited, resulting from a degeneracy splitting of the propagation constants of the modes.
\end{abstract}


\maketitle


The angular momentum of electrons, photons and other quantum particles can be decomposed into spin angular momentum (SAM) and orbital angular momentum (OAM). Spin angular momentum, or polarization for photons, is \emph{intrinsic}, \emph{i.e.} independent of the chosen rotation axis. OAM can be decomposed into \emph{intrinsic} OAM, or IOAM, and \emph{extrinsic} OAM, or EOAM \cite{O'Neil2002, Berry1998, Bliokh2015}. EOAM is associated with the trajectory of the centroid of a wave packet, and is relative to a chosen spatial axis, while IOAM does not depend on axis location, provided that the axis is oriented such that net transverse momentum is zero, as shown by Berry \cite{Berry1998}. For example, the OAM of an electronic energy eigenstate in an atom is intrinsic, as is that of a helical phase vortex within an electron or photon beam \cite{McMorran2011,Allen1992}. In contrast, EOAM exists when a photon travels in a curved path defined by a helically coiled optical fiber. Here we present experimental evidence for the interaction of IOAM with SAM for photons propagating in a straight few-mode cylindrically symmetric waveguide.

Spin-orbit interaction (SOI) involves interaction between SAM and OAM. An example of spin-IOAM interaction is Russell-Saunders \( \vec L\cdot \vec S\)  coupling in a single-electron atom, which splits the degeneracy of electronic energy levels, forming fine structure. An example of spin-EOAM interaction is seen in the precession of the linear polarization vector of photons traveling in a coiled single-mode optical fiber, wherein the photons are forced to follow a three-dimensional path \cite{Tomita1986}. Spin-orbit interactions are deeply connected to a geometric (Berry) phase or gauge potential description, as shown for the intrinsic electron case by Mathur \cite{Mathur1991} and for the extrinsic photon case by Wu and Chiao \cite{Chiao1986}, and summarized by Bliokh \emph{et al.} \cite{Bliokh2015}. 

In the case of a narrow collimated light beam guided along a helical trajectory by many internal reflections in a large glass cylinder \cite{Bliokh2008}, the situation looks analogous to a helically coiled fiber where the light has EOAM. From a different perspective, such a beam can also be described by superpositions of many eigenmodes of the cylinder each of which carries IOAM. This highlights the contextuality of whether OAM is considered intrinsic or extrinsic. Concentrating on a small region of a beam with IOAM, \emph{e.g.} by passing it through an off-center aperture such that the apertured field has net transverse momentum, produces a beam with EOAM \cite{O'Neil2002}. Nevertheless, viewing a beam as a whole leads to a distinction between EOAM and IOAM, as pointed out by Berry \cite{Berry1998}.

In optical fiber made of isotropic material and directed along a straight-line path, interaction between SAM and OAM is mediated by the confining refractive-index gradient through a spin-Hall effect called the optical Magnus effect \cite{Liberman1992}. The refractive-index gradient plays a role analogous to the electric potential gradient's role in the atomic case. Conservation of light's angular momentum upon reflection requires corrections to geometrical optics that couple light's polarization to the trajectory it traverses (and \emph{vice versa}). This is illustrated simply in a ray picture at a sharp interface by the Imbert-Fedorov shift, in which the centroid of a reflected circularly-polarized beam is displaced perpendicular to the plane of reflection in a direction dictated by the polarization handedness \cite{Bliokh2015, Bliokh2013}. Trajectories with OAM do not pass through the fiber axis, so this displacement increases the longitudinal distance between reflections for one polarization while decreasing it for the other. Such shifts are typically subwavelength, but the many reflections light undergoes while traveling in optical fiber can amplify the total effect size up to a macroscopic level.

In a straight highly multimode fiber, (with core diameter much greater than the wavelength of guided light), a speckle pattern is created when many modes interfere coherently. Spin-orbit interaction gives rise to fiber-length dependent rotation of speckle patterns around the fiber axis with a positive or negative direction of rotation determined by the handedness of the circular polarization (helicity) of the light \cite{Dooghin1992}. This phenomenon can be adequately modeled using a ray-tracing or trajectory approximation, highlighting its close connection with EOAM \cite{Liberman1992}.

In the few-mode regime where the core diameter and guided wavelength are similar, diffraction effects become important and a wave picture is preferable. In the wave picture, trajectories are replaced conceptually by mode distributions describing OAM. Spin-IOAM interaction splits the degeneracy of the propagation constants (phase velocities), distinguished by parallel or anti-parallel orientation of spin and intrinsic OAM. The shift due to the Magnus effect is along the direction of energy flow for parallel modes, and opposes the direction of energy flow for anti-parallel modes. Superpositions of split modes with differing phase velocities manifest rotational beating effects, which take their cleanest forms as continual rotation of either spatial-pattern or linear-polarization orientation along the length of the fiber.

Intermodal coupling can complicate observation of such beating effects. In dispersion-tailored fiber, intermodal coupling can be suppressed to allow for stable mode propagation  \cite{RamachandranOVs, Ung2014, Gregg2016}. In this letter, we utilize dispersion-tailored few-mode fiber to measure the interaction between spin and intrinsic optical OAM. Our ability to excite selectively the four modes that have 2 units of IOAM allows clean observation of the resulting rotational beating effects. A recent theory makes the following predictions about the relation between spatial and polarization rotations \emph{vs.} fiber length \cite{Leary2009}: 1) The rotation angles should be linear with fiber length, 2) the spatial rotation rate should be an integer multiple of the polarization rotation rate, depending on the value of the IOAM, 3) the spatial rotation rate should be equal in magnitude and opposite in direction for left- and right-handed circular polarizations, and 4) for a given IOAM value, $|\ell| >1$, the polarization rotation rate should be equal in magnitude and opposite in direction for left- and right-handed IOAM. We present results of an experiment that confirms all four of these predictions, providing strong evidence for the existence of purely intrinsic SOI of light. We focus on the experimentally accessible photon case, but the same model is expected to apply to electron SOI in analogous waveguides \cite{Leary2009}.

Spin-orbit interaction of a photon propagating in a weakly-guiding cylindrically-symmetric waveguide, labeled with cylindrical coordinates $(\rho, \phi, z)$ and time $t$, is described by a time-independent Schr\"odinger-like wave equation, which follows from Maxwell's equations and has eigenvalue $\beta^2$ \cite{Leary2009,Leary2014} 
\begin{align}
& \left( \hat{H}_0 + \hat{H}_{SO} \right) \Psi = \beta^2 \Psi \label{eq:waveEqn} \\
&\hat{H}_0 = \nabla^2_T + k^2(\rho) \\
&\hat H_{SO} = \frac{1}{2 \rho} \frac{\partial V(\rho)}{\partial \rho} \hat{L}_z \hat{S}_z \label{eq:Hso}
\end{align}
with Hamiltonian-like operators $\hat{H}_0$ and $\hat{H}_{SO}$, transverse Laplacian $\nabla^2_T$, $k^2(\rho) = \left( n(\rho) \, \omega/c \right)^2$, where $c$ is the speed of light in vacuum and $n(\rho)$ is the refractive index profile, effective potential $V(\rho) = \ln \left[ n^2(\rho)\right]$, dimensionless z-component spin operator $\hat{S}_z$, dimensionless z-component IOAM operator $\hat{L}_z$, longitudinal propagation constant $\beta$, and wavefunction $\Psi = \Psi (\rho, \phi) \,  \exp \left[{i (\beta z - \omega t)}\right]$ where the angular frequency is $\omega$.

The unperturbed modes of the waveguide are constructed in an eigenbasis of IOAM and SAM by neglecting $\hat{H}_{SO}$ and solving $\hat{H}_0 \Psi^{(0)} = \beta_0^2 \Psi^{(0)}$. Let eigenstates of the spin operator obey $\hat{S}_z \ket{s \sigma} = s \sigma \ket{s \sigma}$. The spin handedness, or helicity, is $\sigma = \pm 1$ and $s=1$ for photons. We call $\sigma = +1$ left-circularly polarized (LCP) and $\sigma=-1$ right-circularly polarized (RCP).  Let IOAM eigenstates obey $\hat{L}_z \ket{\ell} = \ell \ket{\ell}$, with IOAM z-component operator $\hat{L}_z = -i \partial_\phi$. The IOAM eigenvalue of $\hat{L}_z$ is $\ell = \mu |\ell|$, which has handedness $\mu = \pm 1$. Our fiber modes are well modeled with paraxial light under the weak-guidance approximation, where SAM and IOAM are separable \cite{vanEnk1994} (contrast with \cite{Golowich2014}), and the monochromatic bound modes of the waveguide are
\begin{align}\label{eq:modes}
\Psi^{(0)}_{\ell,m} &= F \psi_{\ell,m}^{(0)} e^{i \left( \beta_0 z - \omega t \right)} \hat{e}_\sigma \nonumber \\
&= F \varphi_{\ell, m} (\rho) e^{i  \ell \phi} e^{i \left( \beta_0 z - \omega t \right)} \, \hat{e}_\sigma
\end{align}
where  $\psi^{(0)}_{\ell,m} = \varphi_{\ell, m} (\rho) e^{i  \ell \phi}$ is the transverse spatial distribution, $\varphi_{\ell,m} (\rho)$ is the radial wave function with radial quantum number $m$, $\hat{e}_\sigma$ is the unit circular polarization vector, and $F$ is a normalization constant. For given $\omega$, an unperturbed mode $\Psi^{(0)}_{\ell,m}$ has a propagation constant $\beta_0$ that is degenerate in the signs of $\ell$ and $\sigma$.

\begin{figure}[tb]
\includegraphics[width=0.45\textwidth]{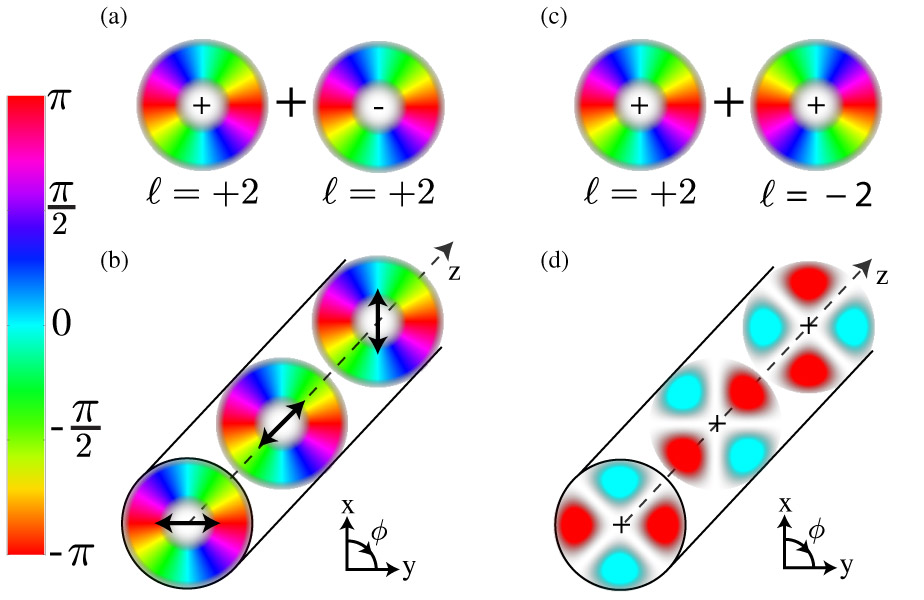}
\caption{\label{fig:ModeCombo}  Superpositions of modes for observing change of propagation constant $\delta \beta$. Color in the legend shows phase. (a) Same IOAM $\ell$ and opposite circular polarizations (+ and -) combine to make (b) rotating linear polarization in \emph{orbit-controlled spin rotation}. (c) Same circular polarizations, but opposite IOAM combine to make (d) a rotating four-lobed spatial pattern in \emph{spin-controlled orbital rotation}.}
\end{figure}

As in atomic spin-orbit interaction, this degeneracy is lifted by perturbative correction. The first-order correction to the propagation constant squared is \(\delta(\beta^2) = \braket{ \Psi^{(0)} | \hat{H}_{SO} | \Psi^{(0)}} \). Let the first-order corrected propagation constant be $\beta_1$, define \(\beta_{1} = \beta_0 + \delta \beta \), and note that $\delta(\beta^2) = \beta_1^2 - \beta_0^2 \approx 2 \beta_0 \delta \beta,$ neglecting $(\delta \beta)^2$ terms. The linearized first-order perturbative correction to the propagation constant is then \cite{Leary2009}
\begin{equation}
\delta \beta = \frac{ s \sigma \ell}{2 \beta_0 N} \int_0^\infty \, \varphi_{\ell,m}^* (\rho) \frac{\partial V(\rho)}{\partial \rho}  \varphi_{\ell, m }(\rho) \diff \rho \label{eq:db}
\end{equation}
where $N=\int_0^\infty \rho  \left| \varphi_{\ell, m }(\rho)\right|^2 \diff \rho$ is for normalization. For fibers, the integral in Eq. \ref{eq:db} is known as the polarization correction integral \cite{[][{, see chapter 5.2.3.}]Bures,[][{, see p. 286, Eq. 13-11.}]SnydLove}. The sign of the splitting $\delta \beta$ is controlled by the product $\sigma \mu$.  We call modes with $\sigma \mu = +1$ parallel modes, as IOAM and SAM are co-oriented, while $\sigma \mu = -1$ are anti-parallel modes. An important caveat occurs specifically in the case of $|\ell|=1$, (and is conspicuously absent in the electron case where $s=1/2$). The two anti-parallel combinations with $\ell + \sigma = 0$, corresponding to transverse electric and transverse magnetic modes, have distinct $\delta \beta$ values, which complicate the effects observed when mode superpositions are excited \cite{Bures,Butkovskaya1997,Darsht1995}. Therefore, we study the case $|\ell| = 2$. For $|\ell| > 1$, Eq. \ref{eq:db} predicts that parallel and anti-parallel modes have propagation constants that differ by $2 \, \delta \beta$ as a result of spin-IOAM interaction. While the effects of spin-EOAM interaction are avoided by using a waveguide along a straight path as we employ here, spin-IOAM interaction is inescapable for $|\ell| > 0$ modes.

 \begin{figure*}
 \includegraphics[width=\textwidth]{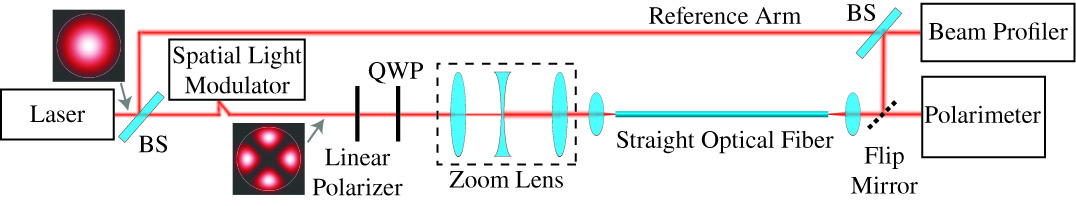}
 \caption{\label{fig:setup}  Experimental apparatus. The Ti:Saph laser is configured for continuous-wave operation. BS = Non-polarizing beamsplitter. QWP = quarter-wave plate. The spatial light modulator converts a Gaussian beam to the desired profile. Inset profiles are simulated. In measuring fiber output beam, the reference arm is blocked to measure the profile and unblocked to measure interferogram.}
 \end{figure*}
Propagation constant splitting implies that parallel and anti-parallel modes accumulate phase at different rates as a function of distance. Thus, \emph{coherent superpositions of parallel and anti-parallel modes exhibit rotational beating as a function of longitudinal propagated distance $z$}, but are stationary with monochromatic excitation and fixed distance. There are two such beating effects (both illustrated in Fig.\ \ref{fig:ModeCombo}) that occur within the fiber and allow for independent measurements of the splitting $\delta \beta$, \cite{Leary2009}. In each case, the sign of one property, called the control property, breaks the symmetry and sets the direction of the rotation associated with the other property. Let $\psi^{(1)}_{\sigma, \ell, m} = \psi^{(0)}_{\ell, m} \exp (-i\sigma  \mu |\delta \beta| z)$ include the propagation constant correction to the mode. One of the two forms of the rotation is \emph{orbit-controlled spin rotation}. Representing the polarization with a Jones vector in a Cartesian basis, $ \hat e_\sigma =  \left[ 1 , i \sigma \right]^T$, superposition of modes with the same IOAM but opposite SAM yield
\begin{align}
& \psi^{(1)}_{+,\mu|\ell|, m} \hat{e}_{+} + \psi^{(1)}_{-,\mu|\ell|, m} \hat{e}_{-} \nonumber \\
=&\varphi_{\ell,m} (\rho)  e^{i  \ell \phi }  \left(  e^{-i \,\mu  |\delta \beta| \,z} \hat{e}_{+} +  e^{i \, \mu|\delta \beta| \,z} \hat{e}_{-}  \right) \nonumber \\
=& 2 \varphi_{\ell,m} (\rho)  e^{i \ell\phi }  \left[ \begin{array}{c} \cos (|\delta \beta| z) \\   \mu \sin (|\delta \beta| z) \end{array} \right]
\end{align}
where the spatial profile is unchanged and the linear polarization rotates with $z$, in a direction controlled by the IOAM handedness $\mu$, by an angle \( \phi = \mu |\delta \beta| z\). The other form of rotation is \emph{spin-controlled orbital rotation}, where superposition of modes with the same SAM but opposite IOAM yield
\begin{align}
& \psi^{(1)}_{\sigma, +|\ell|, m} \hat{e}_\sigma + \psi^{(1)}_{\sigma, -|\ell|, m} \hat{e}_\sigma\nonumber \\
=&\varphi_{\ell,m} (\rho)  \left(  e^{i (|\ell| \,\phi - \sigma \, |\delta \beta| \,z )} +  e^{-i (|\ell| \,\phi -\sigma \, |\delta \beta| \,z )}  \right)  \hat{e}_\sigma \nonumber\\
= &2 \varphi_{\ell,m} (\rho)  \cos \left[|\ell| \left(\phi - \sigma \left|\frac{\delta \beta}{\ell}\right| z \right) \right] \hat{e}_\sigma,
\end{align}
where the polarization remains unchanged while the spatial profile rotates with $z$, in the direction set by $\sigma$, by an angle $\xi = \sigma \left|\frac{\delta \beta}{\ell}\right| z$.

\begin{figure}[b]
\begin{center}
\includegraphics[width=0.48\textwidth]{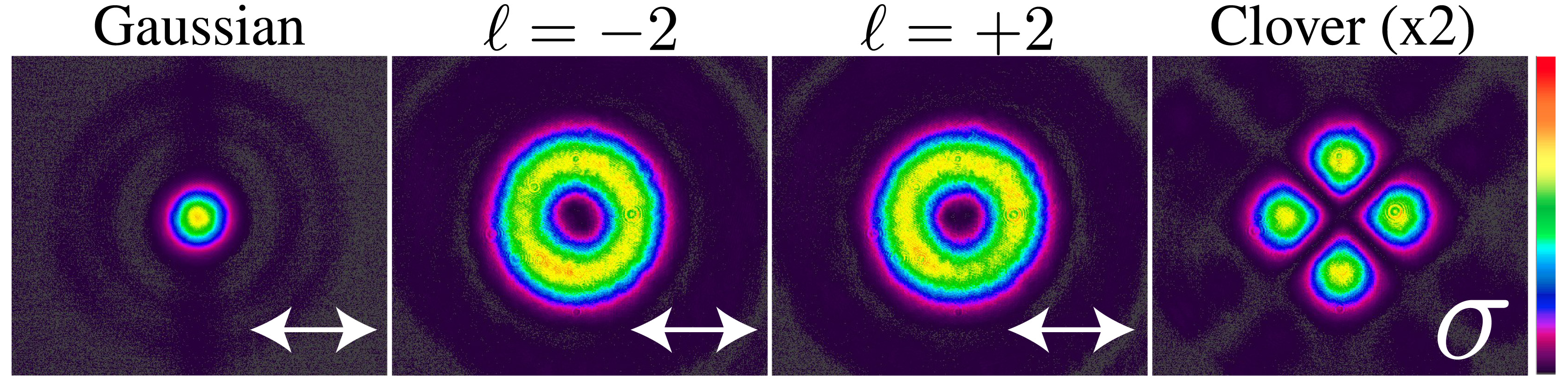}
\caption{ Input profiles. We use a total of five input settings. We probe spin-IOAM interaction with four inputs in the ``experimental group'': two circularly polarized ($\sigma$) clover profiles, and two horizontally polarized (arrowed lines) $\ell = \pm2$ profiles. The ``control group'' consists of horizontally polarized Gaussian profiles, which lack IOAM and are hypothesized to propagate unchanged through the fiber. Right: Intensity color legend.}
\label{fig:input_profiles}
\end{center}
\end{figure}

Cylindrically symmetric optical fiber using the configuration shown in Fig.\ \ref{fig:setup} provides a direct test of this theoretical model. To minimize unwanted coupling between waveguide modes of different order, we use a dispersion-tailored fiber with multiple index steps, in which the $|\ell|=2$ modes have $\beta$ values well separated from those of other modes \cite{Ramachandran2005a}. A Ti:sapphire laser running in a continuous-wave configuration with $\lambda = 799.953$ nm is directed onto a spatial light modulator (SLM) to create the desired transverse spatial profile, which, in turn, excites the desired superpositions of fiber modes with average power of order 100 $\mu W$. We measure the splitting of $|\ell| = 2$ modes using all four combinations that have one parallel and one anti-parallel mode, and call these superpositions the ``experimental group'' input profiles. We also measure one ``control group'' fiber fundamental mode, which has $\ell = 0$ and thus experiences no spin-IOAM interaction. Modes with $|\ell|>2$ are not supported in our experimental fiber at accessible wavelengths. We refer to the profiles with $\cos (2 \phi)$ azimuthal dependence as ``clover'' profiles. Input profiles and interferograms are shown in Fig.\ \ref{fig:input_profiles}. The fork patterns made by the fringes in the $\ell = \pm2$ inferferograms verify the IOAM content of the beams. The SLM holograms and quarter-wave plate control the input profiles and polarization without changing the sensitive optical alignment. The holograms, shown in section 1 of supplementary material (SM) \cite{SM}, produce Laguerre-Gauss spatial modes, which couple well into the exact fiber modes of interest. Clover profiles are superpositions of Laguerre-Gauss modes (see Fig.\ \ref{fig:ModeCombo} (c-d)). Input clover profiles are circularly polarized while all other input profiles are horizontally polarized.
 
The procedure to probe modal evolution along the fiber's longitudinal direction is to cleave short segments ($\sim 1$ cm) off the output end of the fiber and, at each length $L$, excite the input profiles and take output measurements. This approach is complimentary to Wang \emph{et al.}'s investigation utilizing spectroscopic measurement and  fiber Bragg gratings to determine the magnitude of SOI splitting in fiber without directly observing rotation dynamics \cite{Wang2014}. The fiber path must be kept sufficiently straight to avoid rotation due to spin-EOAM interaction (geometric phase rotations, discussed in section 4 of the supplementary material). The fiber is epoxied into place at the input, rests on two platforms topped with double-sided tape as tension relief and the region of the fiber after the tension relief is stripped of its jacket prior to the experiment. After cutting the fiber to a new length, the output is pulled into a position that straightens the fiber, but care is taken to avoid longitudinal strain by pulling no harder than necessary.

\begin{figure}[b]
\begin{center}
\includegraphics[width=0.45\textwidth]{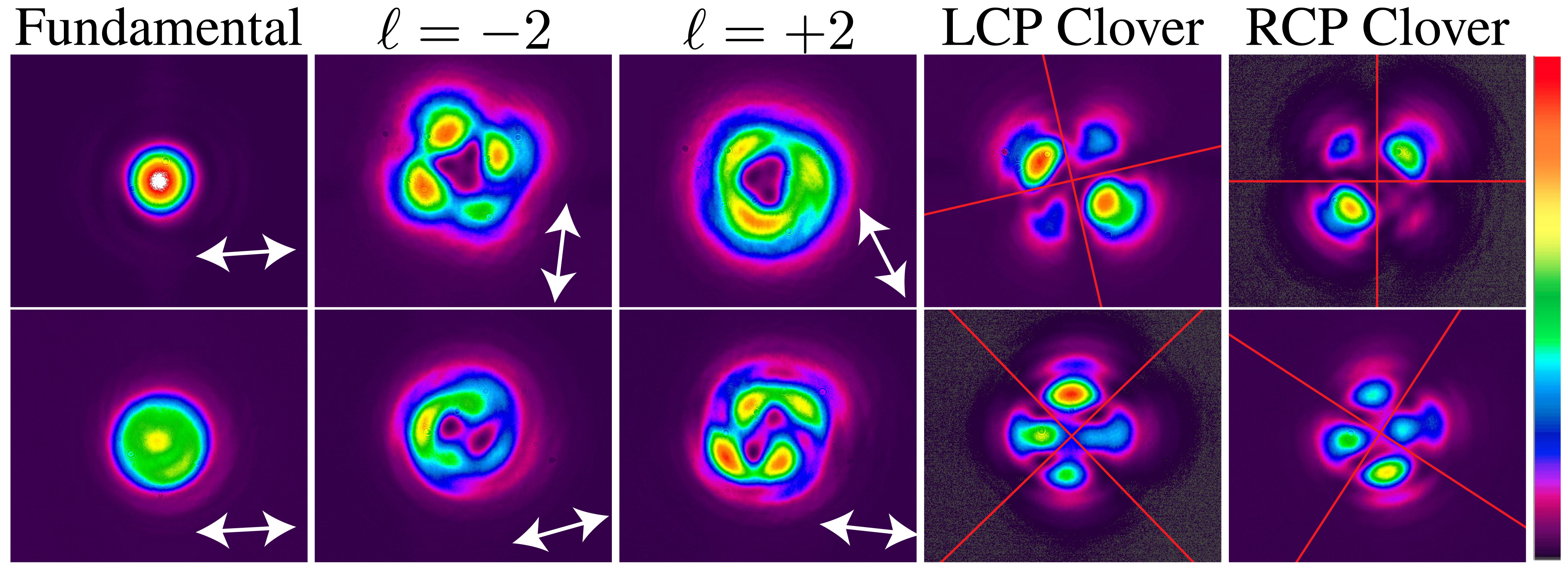}
\caption{ Representative output profile pictures. Top row profiles are at $L=43.5$ cm. Bottom row profiles are at $L=38.4$ cm. Columns labeled with $\ell$ have IOAM input profiles, while LCP and RCP label the circular polarization of clover input profiles. Red crosshairs on clover profiles indicate orientation of nodal lines for $\xi$ measurement, and white arrows indicate the major axis orientation $\phi$ for linearly polarized modes. Width differences due to slight difference in output objective distance from fiber output at different lengths. Right: Intensity color legend.}
\label{fig:profiles}
\end{center}
\end{figure}

\begin{figure}[t]
\includegraphics[width=0.45\textwidth]{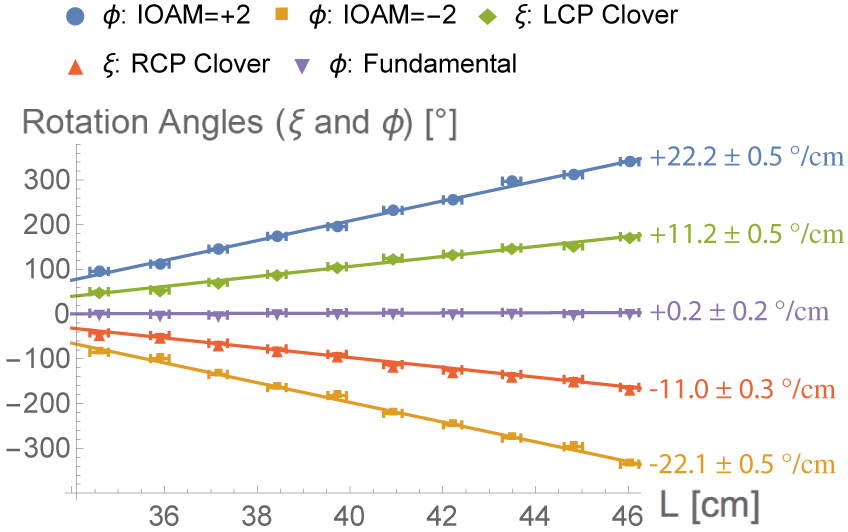}
\caption{\label{fig:measurements}  Measured output parameters $\xi$ (spatial orientation of the clover profiles) and $\phi$ (linear polarization orientations for all non-clover inputs). Vertical error bars are smaller than the data symbols. Both $\xi$ and $\phi$ are in configuration space $^\circ$. Best linear fits shown as colored lines with slopes given in $^\circ$/cm. For ease of comparison, these lines are offset from the lines of total accumulated rotation by integer multiples of $360^\circ$ and cross near 30.8 cm.}
 \end{figure}

We report whole-beam polarization measurements as angles on a Poincar\'e sphere, where $\phi$ is the orientation of the major axis, $\theta$ indicates the ellipticity of the polarization, and the degree of polarization indicates how uniform the polarization is \cite{Saleh}. Section 1 of the SM discusses polarization measurements in more detail \cite{SM}.  Output spatial profiles (see Fig.\ \ref{fig:profiles} for a summary and section 1 of the SM for all data) are recorded on a beam profiler and combined with a reference beam that has a flat phase profile to probe the output beam's phase structure. Orientation angles $\xi$ of clover nodal lines are measured manually by rotating a crosshair along the lines in software. The linearly polarized inputs stay well linearly polarized $(\theta \sim 90^\circ)$. Measured output orientations ($\phi$ and $\xi$) and fit lines are shown in Fig.\ \ref{fig:measurements}.

Residual variations in $\theta$ and the degree of polarization, as well as spatial profile distortion and slight oscillation of the $\phi$ values of $|\ell|=2$ modes are observed (see section 2 of the SM). We believe these all result from weak intermodal coupling. Defect-mediated mode coupling favors energy transfer between modes with similar $\beta$ values \cite{Gregg2015}, so we expect coupling to be dominantly between $|\ell|=2$ modes, which is supported by the retention of the characteristic number ($|\ell|=2$) of nodal lines and phase singularities in the output profiles. The retention of linearity in Fig.\ \ref{fig:measurements} further supports that the unwanted coupling is weak, and that spin-IOAM rotations are robust against these unwanted perturbations.

In agreement with the four theoretical predictions in the introduction and as shown in Fig.\ \ref{fig:measurements}: 1) the rotations are linear with fiber length, 2) the slopes of the best fit lines for spatial and polarization rotations differ by a factor of $|\ell|=2$, and 3) and 4) to within experimental uncertainty, the slopes of the spatial and the polarization rotations, are equal in magnitude and opposite in direction for both control property settings (polarization or IOAM handedness). This agreement indicates that, as expected, the parallel modes are degenerate in propagation constant and the anti-parallel modes are degenerate. The average splitting is measured to be $\delta \beta = (22.1 \pm 0.7)^\circ$/cm \cite{SM}. Furthermore, the fundamental input remains horizontally polarized, as predicted above since it carries zero IOAM. This observation rules out confounding rotation effects and supports that the observed rotations are due to spin-IOAM interaction.

Measurement of spin-IOAM interaction characterizes the fine structure of the propagation constant and lays the foundation for investigation of the simultaneous interaction between spin and both EOAM and IOAM, towards precision encoding of information in the spatial distributions of light in optical fibers \cite{Bozinovic2013}. Identical dynamics are expected in analogous electron waveguide experiments and the present study may motivate such investigations in the future.
 
\begin{acknowledgments}
DLPV was supported by the National Science Foundation (NSF) Grants No. (ECCS-1101811, DGE-0742540). MGR was supported by the NSF Grants No. (ECCS-1101811, PHY-1521466). SR and PG were supported by the Office of Naval Research (ONR) MURI Grant No. N0014-13-1-0672, NSF Grants No. (ECCS-1310493, DGE-1247312), and the Air Force Office of Scientific Research (AFOSR) BRI Grant No. FA9550-14-1-0165. The authors acknowledge Raghuveer Parthasarathy for use of the spatial light modulator, Steven van Enk, Benjamin McMorran, and Jordan Pierce for helpful discussion, and Sonja Franke-Arnold for the code which we modified to produce holograms. We thank L. Gr\"uner-Nielsen from OFS-Fitel LLC. for manufacturing the optical fiber used in the experiments described in this letter.
\end{acknowledgments}


\end{document}